\documentclass[prd,aps,showpacs,nofootinbib,preprint,eqsecnum]{revtex4}
\usepackage{graphicx,color,amsmath,amsxtra}
\usepackage{epsf}
\usepackage{amssymb}
\usepackage{enumerate}
\usepackage{hhline}
\usepackage{array}
\usepackage{tabularx}

\newcommand{\bear}{\begin{array}}  \newcommand{\eear}{\end{array}}
\newcommand{\bea}{\begin{eqnarray}}  \newcommand{\eea}{\end{eqnarray}}
\newcommand{\beq}{\begin{equation}}  \newcommand{\eeq}{\end{equation}}
\newcommand{\bef}{\begin{figure}}  \newcommand{\eef}{\end{figure}}
\newcommand{\bec}{\begin{center}}  \newcommand{\eec}{\end{center}}

\newcommand{\Eqn}[1]{&\hspace{-0.2em}#1\hspace{-0.2em}&}

\def\Vec#1{\mbox{\boldmath $#1$}}

\def\be{\begin{equation}}
\def\ee{\end{equation}}
\def\bea{\begin{eqnarray}}
\def\eea{\end{eqnarray}}
\def\beq{\begin{eqnarray}}
\def\eeq{\end{eqnarray}}

\def\nn{\nonumber \\}
\def\e{{\rm e}}

\def\be{\begin{equation}}
\def\ee{\end{equation}}
\def\bea{\begin{eqnarray}}
\def\eea{\end{eqnarray}}
\def\beq{\begin{eqnarray}}
\def\eeq{\end{eqnarray}}

\def\nn{\nonumber \\}
\def\e{{\rm e}}

\begin{document}

\title{
Crossing of the phantom divide in modified gravity
}

\author{Kazuharu Bamba$^{1,}$\footnote{E-mail address: bamba``at"phys.nthu.edu.tw}, 
Chao-Qiang Geng$^{1,}$\footnote{E-mail address: geng``at"phys.nthu.edu.tw}, 
Shin'ichi Nojiri$^{2,}$\footnote{E-mail address: nojiri``at"phys.nagoya-u.ac.jp} and Sergei D. Odintsov$^{3,}$\footnote{Also at Lab. Fundam. Study, Tomsk State
Pedagogical University, Tomsk. E-mail address: odintsov``at"aliga.ieec.uab.es}}
\affiliation{
$^1$Department of Physics, National Tsing Hua University, Hsinchu, Taiwan 300\\
$^2$Department of Physics, Nagoya University, Nagoya 464-8602, Japan\\
$^3$Instituci\`{o} Catalana de Recerca i Estudis Avan\c{c}ats (ICREA)
and Institut de Ciencies de l'Espai (IEEC-CSIC),
Campus UAB, Facultat de Ciencies, Torre C5-Par-2a pl, E-08193 Bellaterra
(Barcelona), Spain
}


\begin{abstract}
We reconstruct an explicit model of modified gravity in which
a crossing of the phantom divide can be realized.
It is shown that the (finite-time) Big Rip singularity appears in the model
of modified gravity (i.e., in the so-called Jordan frame),
whereas that in the corresponding scalar field theory
obtained through the conformal transformation (i.e., in the so-called
Einstein frame) the singularity becomes the infinite-time one.
Furthermore, we investigate the relations between the scalar field theories 
with realizing a crossing of the phantom divide and the corresponding 
modified gravitational theories by using the inverse conformal transformation.
It is demonstrated that the scalar field theories describing the non-phantom
phase (phantom one with the Big Rip)
can be represented as the theories of real (complex) $F(R)$ gravity through
the inverse (complex) conformal transformation.
We also study a viable model of modified gravity in which
the transition from the de Sitter universe to the phantom phase can occur. 
In addition, we explore the stability for the obtained solutions of the 
crossing of the phantom divide under a quantum correction coming from 
conformal anomaly. 
\end{abstract}


\maketitle

\section{Introduction}

It is observationally confirmed that the current expansion of the universe
is accelerating~\cite{WMAP1, Komatsu:2008hk, SN1}.
Various scenarios to explain the current accelerated expansion of
the universe have been proposed. The mechanism, however, is not well
understood yet (for recent reviews, see~\cite{Peebles:2002gy,
Padmanabhan:2002ji, Copeland:2006wr, review, Nojiri:2008nk}).

There are two approaches to account for the current accelerated expansion of
the universe. One is to introduce some unknown matter, which is called
``dark energy'' in the framework of general relativity.
The other is to modify the gravitational theory, e.g., in simplest
case to study the action described by an arbitrary function of the scalar
curvature $R$, which is called ``$F(R)$ gravity''. Here,
$F(R)$ is an arbitrary function of the scalar curvature $R$ (for reviews,
see~\cite{review, Nojiri:2008nk}).

According to the recent various observational data including the Type Ia
supernovae Gold dataset~\cite{observational status},
there exists the possibility that the effective equation of state (EoS)
parameter, which is the ratio of the effective pressure of the universe to
the effective energy density of it,
evolves from larger than $-1$ (non-phantom phase) to less than $-1$
(phantom one, in which superacceleration is realized; e.g.,
see~\cite{Caldwell:1999ew}), namely,
crosses $-1$ (the phantom divide) currently or in near future.

A number of attempts to realize the crossing of the phantom divide have been
made in the framework of general relativity:
For instance,
scalar-tensor theories with the non-minimal gravitational coupling
between a scalar field and the scalar curvature~\cite{scalar-tensor theories}
or that between a scalar field and the Gauss-Bonnet
term~\cite{scalar-Gauss-Bonnet},
one scalar field model with non-linear kinetic terms~\cite{Vikman:2004dc}
or a non-linear higher-derivative one~\cite{Li:2005fm},
phantom coupled to dark matter with an appropriate
coupling~\cite{Nojiri:2005sx},
the thermodynamical inhomogeneous dark energy model~\cite{Nojiri:2005sr},
multiple kinetic k-essence~\cite{multiple kinetic k-essence},
multi-field models (two scalar fields model~\cite{eli, Caldwell:2005ai,
Wei:2005si}, ``quintom'' consisting of phantom and canonical scalar
fields~\cite{quintom}), and the description of those models through
the Parameterized Post-Friedmann approach~\cite{Fang:2008sn},
or a classical Dirac field~\cite{Cataldo:2007vt}
or
string-inspired models~\cite{string-inspired models},
non-local gravity~\cite{Jhingan:2008ym, Deser:2007jk},
a model in loop quantum cosmology~\cite{Singh:2005km}
and a general consideration of the crossing of the phantom
divide~\cite{Stefancic:2004kb, Kunz:2006wc, Andrianov:2005tm}
(for a detailed review, see~\cite{Copeland:2006wr}).
In fact, however, explicit models of modified gravity realizing the
crossing of the phantom divide have hardly been investigated, although there
were suggestive and interesting related works~\cite{review, abdalla,
Amendola:2007nt}.

In the present paper, we study a crossing of the phantom divide
in modified gravity. We reconstruct an explicit model of modified
gravity in which a crossing of the phantom divide can be realized by using the
reconstruction method proposed in Ref.~\cite{Nojiri:2006gh}.
Furthermore, we investigate the corresponding scalar field 
theory, in which there exist the Einstein-Hilbert action and a scalar field, 
obtained through a conformal transformation of the modified gravitational
theory, and compare the evolution of the universe in the modified
gravitational theory with that in the corresponding scalar field theory.
It is shown that the (finite-time) Big Rip 
singularity~\cite{McInnes:2001zw, Caldwell:2003vq} 
appears in the reconstructed model of modified gravity (i.e., in the
so-called Jordan frame), whereas that in the corresponding scalar field theory
obtained through the conformal transformation (i.e., in the so-called
Einstein frame) the singularity becomes the infinite-time one.
Moreover, we consider the relations between the scalar field theories with 
realizing a crossing of the phantom divide and the corresponding theories of
modified gravity by using the inverse conformal transformation of scalar
field theories.
It is demonstrated that the scalar field theories describing the non-phantom
phase (phantom one with the Big Rip singularity)
can be represented as the theories of real (complex) $F(R)$ gravity through
the inverse (complex) conformal transformation. 
On the other hand, a very realistic model of modified gravity that evades 
solar-system tests and realizes a viable cosmic expansion in the past has 
recently been proposed in Ref.~\cite{Hu:2007nk} (for some related models, 
see~\cite{acc, acc1}).
In this model, our universe is asymptotically de Sitter space. 
Therefore, we also reconstruct a model of modified gravity in which 
the transition from the de Sitter universe to the phantom phase can occur
in such a viable theory. 
In addition, we explore the stability for the obtained solutions of the 
crossing of the phantom divide under a quantum correction, in particular 
conformal anomaly. 

Our goal in this paper is to show that in principle the crossing of the 
phantom divide can be realized in the framework of modified gravity without 
introducing any extra scalar components with the wrong kinetic sign (phantom). 
We reconstruct such an explicit model of modified gravity. 
By presenting it, it can be illustrated that the crossing of the 
phantom divide can occur in modified gravity as the scalar field theories 
in the framework of general relativity. 
The demonstration in this work can be interpreted as a meaningful 
step to construct a more realistic model of modified gravity, which could 
correctly describe the expansion history of the universe. 

This paper is organized as follows.
In Sec.\ II we explain the reconstruction method of modified gravity
proposed in Ref.~\cite{Nojiri:2006gh}. Using this method, we reconstruct
an explicit model of modified gravity in which 
a crossing of the phantom divide can be realized. In particular, we show that
the Big Rip singularity appears in this modified gravitational theory.
In Sec.\ III we consider the corresponding scalar field theory, which is 
obtained by making the conformal transformation of the modified gravitational
theory with realizing a crossing of the phantom divide. We demonstrate that 
the Big Rip singularity does not appear in the corresponding scalar field
theory.
In Sec.\ IV we investigate the relations between scalar field theories 
and the corresponding modified gravitational ones. 
In Sec.\ V we study the viable model of modified gravity in which
the transition from the de Sitter universe to the phantom phase can occur. 
In Sec.\ VI, we examine the stability for the obtained solutions of the 
phantom crossing under a quantum correction coming from conformal anomaly. 
Finally, some summaries and outlooks are given in Sec.\ VII. 
Detailed derivations and explanations about each section are shown in 
Appendixes A--F. 
We use units in which $k_\mathrm{B} = c = \hbar = 1$ and denote the
gravitational constant $8 \pi G$ by ${\kappa}^2$, so that
${\kappa}^2 \equiv 8\pi/{M_{\mathrm{Pl}}}^2$, where
$M_{\mathrm{Pl}} = G^{-1/2} = 1.2 \times 10^{19}$GeV is the Planck mass. 

\section{Reconstruction of modified gravity}

We investigate modified gravity with realizing a crossing of the phantom 
divide by using the reconstruction method. 
(The equivalence between $F(R)$ gravity and the scalar tensor theory was
explicitly shown in Ref.~\cite{nojiri}. The limited case was
given in Ref.~\cite{Carroll:2003wy}.)

\subsection{Reconstruction method}

First, we briefly review the reconstruction method of modified
gravity proposed in Ref.~\cite{Nojiri:2006gh}.

The action of $F(R)$ gravity with general matter is given by
\begin{eqnarray}
S = \int d^4 x \sqrt{-g} \left[ \frac{F(R)}{2\kappa^2} +
{\mathcal{L}}_{\mathrm{matter}} \right]\,,
\label{eq:2.1}
\end{eqnarray}
where $g$ is the determinant of the metric tensor $g_{\mu\nu}$ and
${\mathcal{L}}_{\mathrm{matter}}$ is the matter Lagrangian.

The action (\ref{eq:2.1}) can be rewritten to the following form
by using proper functions $P(\phi)$ and $Q(\phi)$ of a scalar field $\phi$:
\begin{eqnarray}
S=\int d^4 x \sqrt{-g} \left\{ \frac{1}{2\kappa^2} \left[ P(\phi) R + Q(\phi)
\right] + {\mathcal{L}}_{\mathrm{matter}} \right\}\,.
\label{eq:2.2}
\end{eqnarray}
The scalar field $\phi$ may be regarded as an auxiliary scalar field because
$\phi$ has no kinetic term.
It follows from the action (\ref{eq:2.1}) that the equation of motion of
$\phi$ is given by
\begin{eqnarray}
0=\frac{d P(\phi)}{d \phi} R + \frac{d Q(\phi)}{d \phi}\,,
\label{eq:2.3}
\end{eqnarray}
which may be solved with respect to $\phi$ as $\phi=\phi(R)$.
Substituting $\phi=\phi(R)$ into the action (\ref{eq:2.2}),
we find that the expression of $F(R)$ in the action of $F(R)$ gravity in
Eq.~(\ref{eq:2.1}) is given by
\begin{eqnarray}
F(R) = P(\phi(R)) R + Q(\phi(R))\,.
\label{eq:2.4}
\end{eqnarray}

 From the action (\ref{eq:2.2}), we find that the field equation of modified
gravity is given by
\begin{eqnarray}
\frac{1}{2}g_{\mu \nu} \left[ P(\phi) R + Q(\phi) \right]
-R_{\mu \nu} P(\phi) -g_{\mu \nu} \Box P(\phi) +
{\nabla}_{\mu} {\nabla}_{\nu}P(\phi) + \kappa^2
T^{(\mathrm{matter})}_{\mu \nu} = 0\,,
\label{eq:2.5}
\end{eqnarray}
where ${\nabla}_{\mu}$ is the covariant derivative operator associated with
$g_{\mu \nu}$, $\Box \equiv g^{\mu \nu} {\nabla}_{\mu} {\nabla}_{\nu}$
is the covariant d'Alembertian for a scalar field, and
$T^{(\mathrm{matter})}_{\mu \nu}$ is the contribution to
the matter energy-momentum tensor.

We assume the flat
Friedmann-Robertson-Walker (FRW) space-time with the metric,
\begin{eqnarray}
{ds}^2 = -{dt}^2 + a^2(t)d{\Vec{x}}^2\,,
\label{eq:2.6}
\end{eqnarray}
where $a(t)$ is the scale factor.

In this background, the $(\mu,\nu)=(0,0)$ component and
the trace part of the $(\mu,\nu)=(i,j)$ component of Eq.~(\ref{eq:2.5}),
where $i$ and $j$ run from $1$ to $3$, read
\begin{eqnarray}
-6H^2P(\phi(t)) -Q(\phi(t)) -6H \frac{dP(\phi(t))}{dt} + 2\kappa^2\rho = 0\,,
\label{eq:2.7}
\end{eqnarray}
and
\begin{eqnarray}
2\frac{d^2P(\phi(t))}{dt^2}+4H\frac{dP(\phi(t))}{dt}+
\left(4\dot{H}+6H^2 \right)P(\phi(t)) +Q(\phi(t)) + 2\kappa^2 p = 0\,,
\label{eq:2.8}
\end{eqnarray}
respectively,
where $H=\dot{a}/a$ is the Hubble parameter and a dot denotes a time
derivative, $\dot{~}=\partial/\partial t$.
Here, $\rho$ and $p$ are the sum of the energy density and
pressure of matters with a constant
EoS parameter $w_i$, respectively, where $i$ denotes some component of
the matters.

Eliminating $Q(\phi)$ from Eqs.~(\ref{eq:2.7}) and (\ref{eq:2.8}), we obtain
\begin{eqnarray}
\frac{d^2P(\phi(t))}{dt^2} -H\frac{dP(\phi(t))}{dt} +2\dot{H}P(\phi(t)) +
\kappa^2 \left( \rho + p \right) = 0\,.
\label{eq:2.9}
\end{eqnarray}
We note that the scalar field $\phi$ may be taken as $\phi = t$ because
$\phi$ can be redefined properly.

We now consider that $a(t)$ is described as
\begin{eqnarray}
a(t) = \bar{a} \exp \left( \tilde{g}(t) \right)\,,
\label{eq:2.10}
\end{eqnarray}
where
$\bar{a}$ is a constant and $\tilde{g}(t)$ is a proper function.
In this case, Eq.~(\ref{eq:2.9}) is reduced to
\begin{eqnarray}
&&
\frac{d^2P(\phi)}{d\phi^2} -\frac{d \tilde{g}(\phi)}{d\phi}
\frac{dP(\phi)}{d\phi} +2 \frac{d^2 \tilde{g}(\phi)}{d \phi^2}
P(\phi) \nonumber \\
&& \hspace{10mm}
{}+
\kappa^2 \sum_i \left( 1+w_i \right) \bar{\rho}_i
\bar{a}^{-3\left( 1+w_i \right)} \exp
\left[ -3\left( 1+w_i \right) \tilde{g}(\phi) \right] = 0\,,
\label{eq:2.11}
\end{eqnarray}
where $\bar{\rho}_i$ is a constant and we have used
$H= d \tilde{g}(\phi)/\left(d \phi \right)$.
Moreover, it follows from Eq.~(\ref{eq:2.7}) that $Q(\phi)$ is given by
\begin{eqnarray}
Q(\phi) \Eqn{=} -6 \left[ \frac{d \tilde{g}(\phi)}{d\phi} \right]^2 P(\phi)
-6\frac{d \tilde{g}(\phi)}{d\phi} \frac{dP(\phi)}{d\phi} \nonumber \\
&& \hspace{10mm}
{}+
2\kappa^2 \sum_i \bar{\rho}_i \bar{a}^{-3\left( 1+w_i \right)}
\exp
\left[ -3\left( 1+w_i \right) \tilde{g}(\phi) \right]\,.
\label{eq:2.12}
\end{eqnarray}
Hence, if we obtain the solution of Eq.~(\ref{eq:2.11}) with respect to
$P(\phi)$, then we can find $Q(\phi)$. 
In Appendix~A, some points on the reconstruction method are noted. 

We mention that the convenient reconstruction for scalar field theories 
could be given in Refs.~\cite{Nojiri:2005pu, Capozziello:2005tf} 
(for a recent review, see~\cite{sahni}). 
Furthermore, the reconstruction in the scalar-Einstein-Gauss-Bonnet theories 
was considered in Ref.~\cite{Nojiri:2006je}.

\subsection{Explicit model with realizing a crossing of the phantom divide}

Next, using the reconstruction method explained in the preceding subsection,
we reconstruct an explicit model in which 
a crossing of the phantom divide can be realized.

A solution of Eq.~(\ref{eq:2.11}) without matter can be given by 
\begin{eqnarray}
P(\phi) \Eqn{=} e^{\tilde{g}(\phi)/2} \tilde{p}(\phi)\,, 
\label{PDF2} \\
\tilde{g}(\phi) \Eqn{=} - 10 \ln \left[ \left(\frac{\phi}{t_0}\right)^{-\gamma}
 - C \left(\frac{\phi}{t_0}\right)^{\gamma+1} \right]\,, 
\label{PDF4} \\ 
\tilde{p}(\phi) \Eqn{=} \tilde{p}_+ \phi^{\beta_+} + 
\tilde{p}_- \phi^{\beta_-}\,, 
\label{PDF6} \\
\beta_\pm \Eqn{=} \frac{1 \pm \sqrt{1 + 100 \gamma (\gamma + 1)}}{2}\,,
\label{PDF7}
\end{eqnarray}
where $\gamma$ and $C$ are positive constants, 
$t_0$ is the present time, and $\tilde{p}_\pm$ are arbitrary constants. 
The derivation of this solution is shown in Appendix~B. 

 From Eq.~(\ref{PDF4}), we find that $\tilde{g}(\phi)$ diverges at
finite $\phi$ when
\be
\label{PDF8}
\phi = t_s \equiv t_0 C^{-1/(2\gamma + 1)}\ ,
\ee
which tells that there could be the Big Rip singularity at
$t=t_s$~\cite{McInnes:2001zw, Caldwell:2003vq}.
(Other kinds of finite-time future singularities have been studied in
Ref.~\cite{sudden}.)
One only needs to consider the period $0<t<t_s$ because
$\tilde{g}(\phi)$ should be real number.
Eq.~(\ref{PDF4}) also gives the following Hubble rate $H(t)$:
\be
\label{PDF9}
H(t)= \frac{d \tilde{g}(\phi)}{d \phi}
= \left(\frac{10}{t_0}\right) \left[ \frac{ \gamma \left(\frac{\phi}{t_0}\right)^{-\gamma-1 }
 + (\gamma+1) C \left(\frac{\phi}{t_0}\right)^{\gamma} }{\left(\frac{\phi}{t_0}\right)^{-\gamma}
 - C \left(\frac{\phi}{t_0}\right)^{\gamma+1}}\right]\ ,
\ee
where it is taken $\phi=t$.

In the FRW background (\ref{eq:2.6}), even for modified gravity described by
the action (\ref{eq:2.1}), the effective energy-density
and pressure of the universe are given by
$\rho_\mathrm{eff} = 3H^2/\kappa^2$ and
$p_\mathrm{eff} = -\left(2\dot{H} + 3H^2 \right)/\kappa^2$, respectively.
The effective EoS parameter
$w_\mathrm{eff} = p_\mathrm{eff}/\rho_\mathrm{eff}$
is defined as~\cite{review}
\begin{eqnarray}
w_\mathrm{eff} \equiv -1 -\frac{2\dot{H}}{3H^2}\,.
\label{eq:2.16}
\end{eqnarray}
For the case of $H(t)$ in Eq.~(\ref{PDF9}), from Eq.~(\ref{eq:2.16}) we
find that $w_\mathrm{eff}$ is expressed as
\begin{eqnarray}
w_\mathrm{eff} = -1 + U(t)\,,
\label{eq:I0-1-1}
\end{eqnarray}
where
\begin{eqnarray}
U(t) \equiv -\frac{2\dot{H}}{3H^2} =
- \frac{-\gamma + 4\gamma \left( \gamma+1 \right)
\left( \frac{t}{t_s} \right)^{2\gamma+1} +
\left( \gamma+1 \right) \left( \frac{t}{t_s}
\right)^{2\left( 2\gamma+1 \right)} }
{15 \left[ \gamma + \left( \gamma+1 \right)
\left( \frac{t}{t_s} \right)^{2\gamma+1} \right]^2}\,.
\label{eq:I0-1-2}
\end{eqnarray}
Moreover, the scalar curvature is given by
$R=6\left( \dot{H} + 2H^2 \right)$. For the case of Eq.~(\ref{PDF9}),
$R$ is described as
\begin{eqnarray}
R \Eqn{=}
\frac{60
\Biggl[
\gamma \left( 20\gamma -1 \right) + 44\gamma \left( \gamma+1 \right)
\left( \frac{t}{t_s} \right)^{2\gamma+1}
+ \left( \gamma+1 \right) \left( 20\gamma+21 \right)
\left( \frac{t}{t_s} \right)^{2\left( 2\gamma+1 \right)}
\Biggr]
}
{t^2 \left[ 1- \left( \frac{t}{t_s} \right)^{2\gamma+1} \right]^2}
\,.
\label{eq:I0-2}
\end{eqnarray}
In deriving Eqs.~(\ref{eq:I0-1-2}) and (\ref{eq:I0-2}), we have used
Eq.~(\ref{PDF8}).

When $t\to 0$, i.e., $t \ll t_s$, $H(t)$ behaves as
\be
\label{PDF10}
H(t) \sim \frac{10\gamma}{t}\ .
\ee
In this limit, it follows from Eq.~(\ref{eq:2.16}) that
the effective EoS parameter is given by
\be
\label{PDF11}
w_\mathrm{eff} = -1 + \frac{1}{15\gamma}\ .
\ee
This behavior is
identical with that in the Einstein gravity with matter
whose EoS parameter is greater than $-1$.

On the other hand, when $t\to t_s$, we find
\be
\label{PDF12}
H(t) \sim \frac{10}{t_s - t}\ .
\ee
In this case, the scale factor is given by
$a(t) \sim \bar{a} \left( t_s - t \right)^{-10}$.
When $t\to t_s$, therefore, $a \to \infty$, namely, the Big Rip singularity
appears.
In this limit, the effective EoS parameter is given by
\be
\label{PDF13}
w_\mathrm{eff} = - 1 - \frac{1}{15} = -\frac{16}{15}\ .
\ee
This behavior is identical with the case in which there is a phantom matter
with its EoS parameter being smaller than $-1$.
Thus, we have obtained an explicit model showing a crossing of
the phantom divide.

It follows from Eq.~(\ref{eq:2.16}) that the effective EoS parameter
$w_\mathrm{eff}$ becomes $-1$ when $\dot{H}=0$.
Solving $w_\mathrm{eff} = -1$ with respect to
$t$ by using Eq.~(\ref{eq:I0-1-1}), namely, $U(t)=0$,
we find that the effective EoS parameter crosses the phantom divide at
$t=t_\mathrm{c}$ given by
\begin{eqnarray}
t_\mathrm{c} = t_s \left( -2\gamma +
\sqrt{4\gamma^2 + \frac{\gamma}{\gamma+1}}
\right)^{1/\left( 2\gamma + 1 \right)}\,.
\label{eq:I1}
\end{eqnarray}
 From Eq.~(\ref{eq:I0-1-2}), we see that when $t<t_\mathrm{c}$, $U(t)>0$
because $\gamma >0$.
Moreover, the time derivative of $U(t)$ is given by
\begin{eqnarray}
\frac{d U(t)}{dt} =
-\frac{2\gamma \left( \gamma+1 \right) \left( 2\gamma+1 \right)^2}
{15\left[ \gamma + \left( \gamma+1 \right)
\left( \frac{t}{t_s} \right)^{2\gamma+1} \right]^3}
\left( \frac{1}{t_s} \right)
\left( \frac{t}{t_s} \right)^{2\gamma}
\left[ 1 - \left( \frac{t}{t_s} \right)^{2\gamma+1}
\right]\,.
\label{eq:I1-2}
\end{eqnarray}
Eq.~(\ref{eq:I1-2}) tells that the relation $d U(t)/\left(dt\right) <0$ is
always satisfied because we only consider the period $0<t<t_s$ as mentioned
above.
This means that $U(t)$ decreases monotonously. Thus, the value of $U(t)$
evolves from positive to negative. From Eq.~(\ref{eq:I0-1-1}), we see that
the value of $w_\mathrm{eff}$ crosses $-1$.
Once the universe enters the phantom phase, it stays in this phase,
namely, the value of $w_\mathrm{eff}$ remains less than $-1$, and
finally the Big Rip singularity appears because $U(t)$ decreases
monotonically.
Note that other types of the finite-time future singularities in modified
gravity are possible as demonstrated in Ref.~\cite{bamba}.

It follows from Eqs.~(\ref{PDF2}), (\ref{PDF4}), (\ref{PDF6}) and (\ref{PDF8})
that $P(t)$ is given by
\begin{eqnarray}
P(t) = \left[ \frac{\left( \frac{t}{t_0}  \right)^\gamma}
{1-\left( \frac{t}{t_s}  \right)^{2\gamma+1}} \right]^5
\sum_{j=\pm} \tilde{p}_j t^{\beta_j}\,.
\label{eq:I2}
\end{eqnarray}
Using Eqs.~(\ref{eq:2.12}) and (\ref{eq:I2}), one gets
\begin{eqnarray}
Q(t) = -6H
\left[ \frac{\left( \frac{t}{t_0} \right)^\gamma}
{1-\left( \frac{t}{t_s}  \right)^{2\gamma+1}} \right]^5
\sum_{j=\pm} \left( \frac{3}{2}H + \frac{\beta_j}{t} \right)
\tilde{p}_j t^{\beta_j}\,.
\label{eq:I3}
\end{eqnarray}
If we can solve Eq.~(\ref{eq:I0-2}) with respect to $t$ as $t=t(R)$, 
in principle we can obtain the form of $F(R)$ by using this solution and 
Eqs.~(\ref{eq:2.4}), (\ref{eq:I2}) and (\ref{eq:I3}). 
In fact, however, for the general case it is difficult to solve 
Eq.~(\ref{eq:I0-2}) as $t=t(R)$. Hence, as an solvable example, we show the 
behavior of $t_s^2 F(\tilde{R})$ as a function of $\tilde{R} \equiv t_s^2 R$ 
in Fig.~1 for $\gamma =1/2$, 
$\tilde{p}_+ =-1/t_s^{\beta_+}$, $\tilde{p}_- =0$, 
$\beta_+ = \left(1+2\sqrt{19}\right)/2$ and $t_s =2t_0$. 
The quantities in Fig.~1 are shown in dimensionless quantities. 
The horizontal and vertical axes show $\tilde{R}$ and $t_s^2F$, respectively. 
(Here, $\tilde{R} = t_s^2R = 4R/R_0$, where $R_0$ is the current curvature. 
In deriving this relation, we have used $t_s =2t_0$, $t_0 \approx H_0^{-1}$, 
where $H_0$ is the present Hubble parameter.) From Fig.~1, we see that 
the value of $F(R)$ increases as that of $R$ becomes larger.

\begin{figure}[tbp]
\begin{center}
   \includegraphics{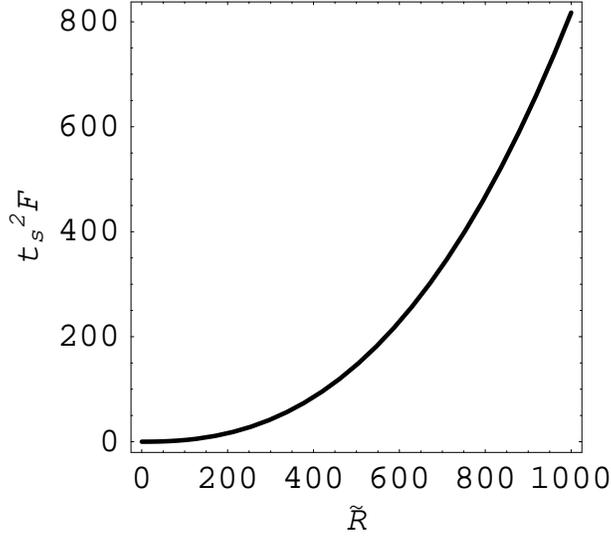}
\caption{Behavior of $t_s^2F(\tilde{R})$ as a function of $\tilde{R}$ for 
$\gamma =1/2$, $\tilde{p}_+ =-1/t_s^{\beta_+}$, 
$\tilde{p}_- =0$, $\beta_+ = \left(1+2\sqrt{19}\right)/2$ and $t_s =2t_0$. 
}
\end{center}
\label{fg:1}
\end{figure}

To examine the analytic form of $F(R)$ for the general case, we investigate 
the behavior of $F(R)$ in the limits $t \to 0$ and $t \to t_s$. 
When $t \to 0$, from Eq.~(\ref{PDF10}) we find 
\begin{eqnarray}
t \sim \sqrt{ \frac{60\gamma \left( 20\gamma -1 \right)}{R} }\,.
\label{eq:I4}
\end{eqnarray}
In this limit, it follows from Eqs.~(\ref{eq:2.4}), (\ref{PDF10}),
(\ref{eq:I2}), (\ref{eq:I3}) and (\ref{eq:I4}) that
the form of $F(R)$ is given by
\begin{eqnarray}
F(R) \Eqn{\sim}
\left\{
\frac{\left[\frac{1}{t_0} \sqrt{60\gamma \left( 20\gamma -1 \right)} R^{-1/2}
\right]^\gamma}{1 -
\left[\frac{1}{t_s} \sqrt{60\gamma \left( 20\gamma -1 \right)} R^{-1/2}
\right]^{2\gamma+1}} \right\}^5 R \nonumber \\
&& \hspace{10mm}
{}\times
\sum_{j=\pm}
\biggl\{ \left( \frac{5\gamma -1 -\beta_j}{20\gamma -1} \right) \tilde{p}_j
\left[60\gamma \left( 20\gamma -1 \right) \right]^{\beta_j /2}
R^{-\beta_j /2}
\biggr\}\,.
\label{eq:I5}
\end{eqnarray}

On the other hand, when $t\to t_s$, from Eq.~(\ref{PDF12}) we obtain
\begin{eqnarray}
t \sim t_s - 3\sqrt{ \frac{140}{R} }\,.
\label{eq:I6}
\end{eqnarray}
In this limit, it follows from Eqs.~(\ref{eq:2.4}), (\ref{PDF12}),
(\ref{eq:I2}), (\ref{eq:I3}) and (\ref{eq:I6}) that
the form of $F(R)$ is given by
\begin{eqnarray}
F(R) \Eqn{\sim}
\left(
\frac{
\left\{ \frac{1}{t_0}
\left[ t_s - 3\sqrt{140} R^{-1/2} \right]
\right\}^\gamma}
{1 -
\left[ 1 - \frac{3\sqrt{140}}{t_s} R^{-1/2}
\right]^{2\gamma+1}
} \right)^5 R
\sum_{j=\pm}
\tilde{p}_j
\left[ t_s - 3\sqrt{140} R^{-1/2}
\right]^{\beta_j}
\nonumber \\
&& \hspace{0mm}
{}\times
\Biggl\{
1- \sqrt{\frac{20}{7}}
\left[
\sqrt{\frac{15}{84}} t_s
+ \left( \beta_j - 15 \right) R^{-1/2}
\right]
\frac{1}{t_s - 3\sqrt{140} R^{-1/2}}
\Biggr\}\,.
\label{eq:I7}
\end{eqnarray}
The above modified gravity may be considered as some approximated form of 
more realistic viable theory. 
For large $R$, namely, $t_s^2R \gg 1$, the expression of $F(R)$ in 
(\ref{eq:I7}) can be approximately written as 
\begin{eqnarray}
F(R) \approx
\frac{2}{7}
\left[
\frac{1}{3\sqrt{140} \left( 2\gamma +1 \right)}
\left( \frac{t_s}{t_0} \right)^\gamma \right]^5
\left(
\sum_{j=\pm}
\tilde{p}_j t_s^{\beta_j} \right)
t_s^5 R^{7/2}\,.
\label{eq:I8}
\end{eqnarray}

\section{Corresponding scalar field theory}

In this section, motivated by the discussion in
Ref.~\cite{Briscese:2006xu},
we consider the corresponding scalar field theory to modified gravity with 
realizing a crossing of the phantom divide, which is obtained
by making the conformal transformation of the modified gravitational
theory.

By introducing two scalar fields $\zeta$ and $\xi$, we can rewrite
the action~(\ref{eq:2.1}) to the following form~\cite{review}:
\begin{eqnarray}
S \Eqn{=}
\int d^{4}x \sqrt{-g}
\left\{
\frac{1}{2\kappa^2} \left[ \xi \left( R-\zeta \right) + F(\zeta) \right] +
{\mathcal{L}}_{\mathrm{matter}}
\right\}\,.
\label{eq:3.1}
\end{eqnarray} 
This is the action in the Jordan-frame, in which there exists a non-minimal
coupling between $\xi$ and the scalar curvature $R$. 
The form in Eq.~(\ref{eq:3.1}) is reduced to the original one in
Eq.~(\ref{eq:2.1}) by using the equation $\zeta=R$,
which is the equation of motion of one auxiliary field $\xi$. 

We make the following conformal transformation:
\begin{eqnarray}
g_{\mu \nu} \hspace{0.5mm} \rightarrow \hspace{0.5mm}
\hat{g}_{\mu \nu} = e^{\sigma} g_{\mu \nu}\,,
\label{eq:3.4}
\end{eqnarray}
where
\begin{eqnarray}
e^{\sigma} = F^{\prime}(\zeta)\,.
\label{eq:3.5}
\end{eqnarray}
Here, $\sigma$ is a scalar field and a hat denotes quantities 
in the Einstein frame, in which the non-minimal coupling between 
$\xi$ and $R$ in the action (\ref{eq:3.1}) disappears. 

By defining $\varphi$ as $\varphi \equiv \sqrt{3/2} \sigma/\kappa$, 
we obtain the following canonical scalar field theory:
\begin{eqnarray}
S_{\mathrm{ST}} =
\int d^{4}x \sqrt{-\hat{g}}
\left[
\frac{\hat{R}}{2\kappa^2} - \frac{1}{2} \hat{g}^{\mu\nu}
{\partial}_{\mu} \varphi {\partial}_{\nu} \varphi
-V(\varphi)
+ e^{-2\sqrt{2/3} \kappa \varphi} {\mathcal{L}}_{\mathrm{matter}}
\right]\,.
\label{eq:3.8}
\end{eqnarray}
The detailed derivation of the action (\ref{eq:3.8}) is given 
in Appendix~C. 

We now investigate the case in which $F(R)$ is given by
\begin{eqnarray}
F(R) = c_1 M^2 \left( \frac{R}{M^2} \right)^{-n}\,,
\label{eq:3.13}
\end{eqnarray}
where
$c_1$ is a dimensionless constant and $M$ denotes a mass scale.
The form of $F(R)$ in Eq.~(\ref{eq:I8}) corresponds to the one in
Eq.~(\ref{eq:3.13}) with $n=-7/2$. 
It may seem that such a model may have problems in the description of the 
past universe evolution. However, there is the trick to make its past 
evolution consistent with observations described in Ref.~\cite{Nojiri:2006gh}. 
It uses the introduction of compensating dark energy dominated at intermediate 
universe which disappears effectively at current universe. 
Our primary purpose in this work is current universe admitting the phantom 
divide crossing in modified gravity, so we will not discuss the past 
evolution of the model under discussion. 
In this case, the scale factor
$a(t)$ and the scalar curvature $R$ are given by~\cite{Briscese:2006xu}
\begin{eqnarray}
a(t) = \bar{a} \left( t_s - t \right)^{(n+1)(2n+1)/(n+2)}\,,
\label{eq:3.14}
\end{eqnarray}
and
\begin{eqnarray}
R \Eqn{=} \frac{6n(n+1)(2n+1)(4n+5)}{(n+2)^2}
\frac{1}{\left( t_s - t \right)^2}\,,
\label{eq:3.15}
\end{eqnarray}
respectively. From Eqs.~(\ref{eq:3.4}) and (\ref{eq:3.5}), we find
\begin{eqnarray}
d\hat{t} \Eqn{=} \pm e^{\sigma/2} dt\,,
\label{eq:3.16} \\
e^{\sigma/2} \Eqn{=} \sqrt{-n c_1}
\left[ \frac{(n+2)^2}{6n(n+1)(2n+1)(4n+5)} \right]^{(n+1)/2}
M^{n+1} \left( t_s - t \right)^{n+1}\,,
\label{eq:3.17}
\end{eqnarray}
where we have used Eq.~(\ref{eq:3.15}).
It follow from Eq.~(\ref{eq:3.17}) that the relation between
the cosmic time in the Einstein frame $\hat{t}$ and that in the
Jordan frame is given by
\begin{eqnarray}
\hat{t} = \mp \frac{\sqrt{-n c_1}}{n+2}
\left[ \frac{(n+2)^2}{6n(n+1)(2n+1)(4n+5)} \right]^{(n+1)/2}
M^{n+1} \left( t_s - t \right)^{n+2}\,.
\label{eq:3.18}
\end{eqnarray}
If $n<-2$, the limit of $t \to t_s$ corresponds to that of
$\hat{t} \to \mp \infty$. For the case of Eq.~(\ref{eq:I8}),
$n =-7/2$. From Eqs.~(\ref{eq:3.4}) and (\ref{eq:3.5}), we also find that
the metric in the Einstein frame is expressed as
\begin{eqnarray}
d\hat{s}^2 = e^\sigma ds^2 = -d\hat{t}^2 + \hat{a}\left( \hat{t} \right)
d{\Vec{x}}^2\,,
\label{eq:3.19}
\end{eqnarray}
where $\hat{a}\left( \hat{t} \right)$ is the scale factor in the scalar field
theory given by
\begin{eqnarray}
\hspace{-7mm}
\hat{a}\left( \hat{t} \right) \Eqn{=} \hat{\bar{a}}
\hat{t}^{3\left[\left(n+1\right)/\left(n+2\right) \right]^2}\,,
\label{eq:3.20} \\
\hspace{-7mm}
\hat{\bar{a}} \Eqn{=}
\bar{a} \left( \mp \frac{1}{n+2}
\right)^{-3\left[\left(n+1\right)/\left(n+2\right) \right]^2}
\nonumber \\
&&
{}\times
\left\{ \sqrt{-n c_1}
\left[ \frac{(n+2)^2}{6n(n+1)(2n+1)(4n+5)} \right]^{(n+1)/2}
M^{n+1} \right\}^{-\left(2n^2 +2n -1 \right)/\left(n+2\right)^2}\,.
\label{eq:3.21}
\end{eqnarray}
For $n =-7/2$, because when $t \to t_s$, $\hat{t} \to \mp \infty$,
it follows from Eq.~(\ref{eq:3.20}) that the scale factor in the scalar field
theory $\hat{a}\left( \hat{t} \right)$ diverges at infinite time.

Consequently, the `finite-time' Big Rip singularity in $F(R)$ gravity, 
i.e., in the Jordan frame, becomes the `infinite-time' one in the 
corresponding scalar field theory obtained through the conformal 
transformation of the theory of $F(R)$ gravity, namely, in the Einstein frame. 
This shows the physical difference of late-time cosmological evolutions
between the theory of $F(R)$ gravity and the corresponding scalar
field theory, which are mathematically equivalent theories.

\section{Relations between scalar field theories and the corresponding 
theories of $F(R)$ gravity}

In this section, following the considerations in
Refs.~\cite{Briscese:2006xu, Capozziello:2005mj},
we investigate the relations between scalar field theories and the 
corresponding transformations to $F(R)$ gravity.

The action of scalar field theories in the Einstein frame is given by 
\begin{eqnarray}
S_{\chi} =
\int d^{4}x \sqrt{-\hat{g}}
\left[
\frac{\hat{R}}{2\kappa^2} \mp \frac{1}{2}
\hat{g}^{\mu\nu} {\partial}_{\mu} \chi {\partial}_{\nu} \chi
-\tilde{W}(\chi) \right]\,. 
\label{eq:4.10}
\end{eqnarray}
Here, in the non-phantom phase the sign of the kinetic term is $-$, 
while in the phantom one that is $+$. 

To study the corresponding theories of $F(R)$ gravity, we make the inverse
conformal transformation of the action of the scalar field
theories~(\ref{eq:4.10}).
In the non-phantom phase, we use the inverse conformal
transformation~\cite{Capozziello:2005mj} in order to 
vanish the kinetic term of $\chi$
\begin{eqnarray}
\hat{g}_{\mu \nu} \hspace{0.5mm} \rightarrow \hspace{0.5mm}
g_{\mu \nu} = e^{\pm \sqrt{2/3} \kappa \chi}
\hat{g}_{\mu \nu}\,.
\label{eq:4.16}
\end{eqnarray}
As a consequence, 
the action in the Jordan frame for the non-phantom phase is given by 
\begin{eqnarray}
S_\mathrm{NP} \Eqn{=}
\int d^{4}x \sqrt{-g}
\frac{F_\mathrm{NP}(R)}{2\kappa^2}\,,
\label{eq:4.19} \\
F_\mathrm{NP}(R) \Eqn{\equiv}
e^{\pm \sqrt{2/3} \kappa \chi(R)} R -
2\kappa^2
e^{\pm 2\sqrt{2/3} \kappa \chi(R)} \tilde{W}\left(\chi(R)\right)\,.
\label{eq:4.20}
\end{eqnarray} 
The scalar field $\chi$ is just an auxiliary field and can be expressed 
in terms of the scalar curvature as $\chi = \chi(R)$ by solving 
the equation of motion of $\chi$: 
\begin{eqnarray}
R= e^{\pm \sqrt{2/3} \kappa \chi}
\left( 4\kappa^2 \tilde{W}(\chi) \pm \sqrt{6} \kappa
\frac{d \tilde{W}(\chi)}{d \chi} \right)\,.
\label{eq:4.18}
\end{eqnarray}

Similarly,
in the phantom phase we use the complex conformal
transformation~\cite{Briscese:2006xu} in order to vanish the 
kinetic term of $\chi$
\begin{eqnarray}
\hat{g}_{\mu \nu} \hspace{0.5mm} \rightarrow \hspace{0.5mm}
g_{\mu \nu} = e^{\pm i \sqrt{2/3} \kappa \chi} \hat{g}_{\mu \nu}\,.
\label{eq:4.21}
\end{eqnarray}
As a result, the action in the Jordan frame for the phantom phase is given by 
\begin{eqnarray}
S_\mathrm{P} \Eqn{=}
\int d^{4}x \sqrt{-g}
\frac{F_\mathrm{P}(R)}{2\kappa^2}\,,
\label{eq:4.24} \\
F_\mathrm{P}(R) \Eqn{\equiv}
e^{\pm i \sqrt{2/3} \kappa \chi(R)}R -
2\kappa^2 e^{\pm i 2\sqrt{2/3} \kappa \chi(R)}
\tilde{W}\left(\chi(R)\right)\,.
\label{eq:4.25}
\end{eqnarray}
The equation of motion of $\chi$ is given by
\begin{eqnarray}
R= e^{\pm i \sqrt{2/3} \kappa \chi}
\left( 4\kappa^2 \tilde{W}(\chi) \mp i \sqrt{6} \kappa
\frac{d \tilde{W}(\chi)}{d \chi} \right)\,.
\label{eq:4.23}
\end{eqnarray}
This equation can be solved with respect to $\chi$ as $\chi = \chi(R)$.

In general, scalar field theories describing the non-phantom (phantom) phase
can be represented as the theories of real (complex) $F(R)$ gravity through
the inverse (complex) conformal
transformation~\cite{Capozziello:2005mj, Briscese:2006xu}.
We note that the consideration of this section can be applied to
not only the model in Eq.~(\ref{PDF9}) but also to any
other scalar field theories with/without the crossing of the phantom
divide: e.g., the case in which the Hubble rate is
given by Eq.~(\ref{eq:I9}). 
In Appendix~D, we examine the more detailed relation between the scalar 
field theories with realizing a crossing of the phantom divide and the 
corresponding theories of $F(R)$ gravity.

\section{Model of $F(R)$ gravity with the transition from the de Sitter
universe to the phantom phase}

In this section, we reconstruct a model of $F(R)$ gravity in which
the transition from the de Sitter universe to the phantom phase can occur
by using the method explained in Sec.~II A.

\subsection{Reconstruction of the viable $F(R)$ gravity}

The interesting viable model is proposed in Ref.~\cite{Hu:2007nk}. 
It is known that the above model is a very realistic modified gravitational 
theory that evade solar-system tests, 
which was mentioned also in Ref.~\cite{Chiba:2003ir}. 
As shown in the above reference~\cite{Hu:2007nk}, this model could 
reproduce the viable cosmic expansion, correctly describing the phases 
before dark energy epoch. 
Hence our universe is asymptotically de Sitter space. 
The form of the model in Ref.~\cite{Hu:2007nk} and its generalization 
is presented in Appendix~E. 

As an example realizing the transition from
the de Sitter universe to the phantom phase,
we can consider the following form of the Hubble rate:
\be
\label{dp1}
H=g_0 + \frac{g_1}{t_s - t}\ ,
\ee
where $g_0$, $g_1$ and $t_s$ are positive constants.
When $t\to -\infty$, $H$ goes to a constant $H\to g_0$. Hence the universe is 
asymptotically de Sitter space.
On the other hand, when $t\to t_s$, the second term on the right-hand side 
of Eq.~(\ref{dp1}) dominates and $H$ behaves as
$H\sim g_1/\left( t_s - t \right)$.
It follows from $\dot{H} \sim g_1/\left( t_s - t \right)^2 >0$
and Eq.~(\ref{PDF11}) that $w_\mathrm{eff}<-1$, namely, the universe
enters the phantom phase. Then there appears the Big Rip singularity
at $t=t_s$.
For the case of Eq.~(\ref{dp1}), $R$ is given by
\begin{eqnarray}
R= 6\left[ 2g_0^2 + \frac{4 g_0 g_1}{t_s-t} +
\frac{g_1 \left(2g_1+1\right)}{\left(t_s-t\right)^2}
\right]\,.
\label{eq:5J-0-1}
\end{eqnarray}

We consider the case in which the contribution from matter could be neglected. 
We take into account it later. 
Eq.~(\ref{dp1}) shows
\be
\label{dp2}
\frac{d \tilde{g}(\phi)}{d\phi} = g_0 + \frac{g_1}{t_s - \phi}\ ,
\ee
where we have taken $\phi=t$.
Substituting Eq.~(\ref{dp2}) into Eq.~(\ref{eq:2.11}), we obtain
\be
\label{dp3}
0 = \frac{d^2 P(\phi)}{d\phi^2} - \left(g_0 + \frac{g_1}{t_s - \phi}\right)
\frac{d P(\phi)}{d\phi}
+\frac{2g_1}{\left( t_s - \phi \right)^2} P(\phi)\,.
\ee
The solution is given by
\be
\label{dp4}
P(z) = C_+ z^\alpha F_K\left(\alpha,\tilde{\gamma}; z\right)
+ C_- z^{1-\tilde{\gamma}}
F_K\left(\alpha - \tilde{\gamma} + 1, 2 - \tilde{\gamma}; z\right)\,,
\ee
where
\bea
\label{dp5}
&& z\equiv g_0\left(\phi - t_s\right)\ , \quad
\alpha\equiv \frac{1 - g_1 \pm \sqrt{g_1^2 - 10g_1 + 1}}{4}\ ,\quad
\tilde{\gamma} \equiv 1\pm \frac{\sqrt{g_1^2 - 10g_1 + 1}}{2}\ ,\nn
&& F_K\left(\alpha,\tilde{\gamma};z\right)=\sum_{n=0}^\infty
\frac{\alpha(\alpha + 1)\cdots (\alpha + n -1)}
{\tilde{\gamma} \left(\tilde{\gamma} + 1\right)\cdots
\left(\tilde{\gamma} + n - 1\right)} \frac{z^n}{n!}\ .
\eea
Here, $F_K$ is the Kummer functions (confluent hypergeometric function),
and $C_+$ and $C_-$ are dimensionless constants.
Using Eqs.~(\ref{eq:2.12}) and (\ref{dp4}), we obtain
\begin{eqnarray}
&&
Q(z) = -6g_0^2 \left(1-\frac{g_1}{z}\right)
\Biggr\{
C_+ z^{\alpha-1} \left[ \left(\alpha -g_1 +z \right)
F_K\left(\alpha, \tilde{\gamma}; z\right) + \frac{\alpha}{\tilde{\gamma}}
zF_K\left(\alpha + 1, \tilde{\gamma}+1; z\right)
\right]
\nonumber \\
&&
\hspace{45mm}
{}+C_- z^{-\tilde{\gamma}}
\biggl[ \left(1-\tilde{\gamma} -g_1 +z \right)
F_K\left(\alpha -\tilde{\gamma} +1, 2 -\tilde{\gamma}; z\right) \nonumber \\
&&
\hspace{45mm}
{}+\frac{\alpha-\tilde{\gamma}+1}{2-\tilde{\gamma}}z
F_K\left(\alpha -\tilde{\gamma} +2, 3 -\tilde{\gamma}; z\right)
\biggr]
\Biggl\}\,.
\label{eq:5J-1}
\end{eqnarray}
In Eqs.~(\ref{dp4}) and (\ref{eq:5J-1}), because we take $\phi =t$,
$z =g_0 \left(t-t_s\right)$.

It follows from Eq.~(\ref{eq:5J-0-1}) that when $t\to -\infty$, namely,
in de Sitter phase, $R$ becomes constant as $R \sim 12 g_0^2$.
On the other hand, when $t\to t_s$,
$R \sim 6g_1 \left(2g_1+1\right)/\left(t_s-t\right)^2$.
Using this relation and $z =g_0 \left(t-t_s\right)$, we find
\begin{eqnarray}
z \sim -g_0 \sqrt{\frac{6g_1 \left(2g_1+1\right)}{R}}\,.
\label{eq:5J-2}
\end{eqnarray}
In the limit of $t\to t_s$, $|z| \ll 1$ because $R$ diverges.
Expanding the Kummer functions in Eqs.~(\ref{dp4}) and (\ref{eq:5J-1})
and taking the first leading order in $z$, from Eqs.~(\ref{eq:2.4}) and
(\ref{eq:5J-2}) we find that the form of $F(R)$ in this limit is
approximately expressed as
\begin{eqnarray}
&&
\hspace{-10mm}
F(R) \approx \frac{R}{\left(\tilde{\gamma}-1\right)\left(2g_1+1\right)}
\biggl\{
C_+ \left(\alpha-1\right) \left(\alpha +g_1 +1\right)
\left[-g_0 \sqrt{6g_1 \left(2g_1+1\right)} \right]^{\alpha} R^{-\alpha/2}
\nonumber \\
&&
\hspace{20mm}
{}-C_- \left(\alpha-\tilde{\gamma}\right) \left(2-\tilde{\gamma}+g_1\right)
\left[-g_0 \sqrt{6g_1 \left(2g_1+1\right)} \right]^{1-\tilde{\gamma}}
R^{-\left(1-\tilde{\gamma}\right)/2}
\biggr\}\,.
\label{eq:5J-3}
\end{eqnarray}

Next, we study the case in which there exists the matter. 
We consider the cold dark matter with $w=0$. 
We numerically solve Eq.~(\ref{eq:2.11}) with the cold dark 
matter. To execute this, for simplicity, we set $t_s g_0 = g_1$ in 
Eq.~(\ref{dp1}). 
We show the behavior of $F(\tilde{R})/\left(2 \kappa^2 \right)$ in Fig.~2. 
From Fig.~2, we see that $F(\tilde{R})$ increases in terms of $\tilde{R}$. 
The detailed explanation of the numerical calculations is given 
in Appendix~F.

\begin{figure}[tbp]
\begin{center}
   \includegraphics{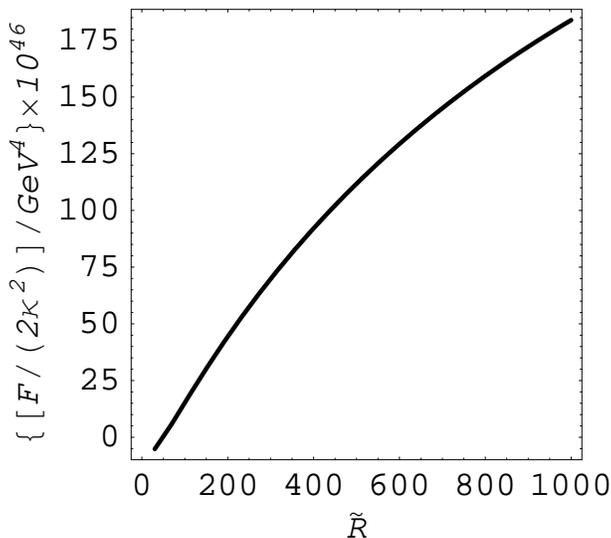}
\caption{
Behavior of $F(\tilde{R})/\left(2 \kappa^2 \right)$ as a function of 
$\tilde{R}$. 
Legend is the same as Fig.~3.
}
\end{center}
\label{fg:2}
\end{figure}

\subsection{Addition of a non-local term to the viable modified gravity
models}

It seems very difficult in the framework of
$F(R)$ gravity to construct a model generating the transition from de Sitter
space to the phantom phase in the viable model of the previous sub-section.
One may add any term, which is a function of $R$, to such models.
The term should be small in the present universe and may be small even in the
past universe, where the curvature could be large. Hence the term should
dominate only at small curvature, that is, smaller one than the present
curvature.
In the asymptotically-de Sitter model above, however, the present universe
is asymptotic de Sitter space,
where the curvature is (almost) constant. Thus, the curvature cannot
become smaller than
the order of the curvature in the present universe and the added term never
dominates.

Let us add a non-local term to any model from Ref.~\cite{acc} by using
a proper function $K$:
\be
\label{df1}
\delta f = K\left(- \Box^{-1} R\right) R^m\ ,
\ee
where $m$ is a positive constant. In the de Sitter universe where the
curvature and the Hubble rate are constant $R=R_0$, $H=H_0$,
we find
\be
\label{df2}
\Box^{-1} R = - \frac{R_0}{3H_0} t + c_4 \e^{-3H_0 t} + c_5\ ,
\ee
where $c_4$ and $c_5$ are constants of the integration. For large $t$, we find
$- \Box^{-1} R \sim \left[ R_0/\left(3H_0\right) \right] t$.
If we choose $K$ to be a slowly
increasing function, $\delta f$ dominates in the future.
If $K$ is  slowly varying
function and could be regarded to be a constant, the total $F(R)$ behaves as
$F(R)\sim R^m$, which gives
\be
\label{df3}
H = \frac{\tilde{h}_0}{t}\ ,\quad \tilde{h}_0 = - \frac{(m-1)(2 m -1)}{m-2}\ ,
\ee
when $\tilde{h}_0>0$ or
\be
\label{df4}
H = \frac{- \tilde{h}_0}{t_s - t}\ ,\quad
\tilde{h}_0 = - \frac{(m-1)(2 m -1)}{m-2}\ ,
\ee
when $\tilde{h}_0<0$. Here, $\tilde{h}_0$ is a constant. 

Alternatively, the addition of an extra scalar field may bring the
evolution to the phantom era.
This is because such terms can become larger at the constant curvature and
hence may induce such a crossing in realistic models.

At present, it is not so clear if such non-local models could be variable or
not due to technical problems, but this model has a possibility to explain the 
complicated cosmic expansion, especially the coincidence problem.
We now consider the models including such non-local terms to show 
the generality of our method.

We remark that as shown in Ref.~\cite{Deser:2007jk}, such a theory may 
successfully pass the solar system tests, 
and that as demonstrated in Ref.~\cite{Jhingan:2008ym}, such non-local models 
may correctly reproduce the whole expansion history of the universe expansion, 
at least in their equivalent scalar-tensor form.

\section{Stability under a quantum correction}

In this section, we examine the stability for the obtained solutions of 
the crossing of the phantom divide under a quantum correction of 
massless conformally-invariant fields. It is convenient to do it by 
taking account of conformal anomaly induced effective pressure and 
energy-density. Note that we do not discuss the quantum regime of modified 
gravity itself because it is relevant at strong curvature (near to the Planck 
scale) where the form of modified gravity may be quite different from the 
one at the late universe. 

Quantum effects produce the conformal anomaly: 
\be
\label{CA1}
T_A=b\left(F+ \frac{2}{3}\Box R\right) + b' G + b''\Box R\ ,
\ee
Here $F$ : the square of 4d Weyl tensor, $G$ : Gauss-Bonnet invariant,
which are given as
\be
\label{CA2}
F = \frac{1}{3}R^2 -2 R_{ij}R^{ij}+ R_{ijkl}R^{ijkl} \ ,\quad 
G = R^2 -4 R_{ij}R^{ij}+ R_{ijkl}R^{ijkl} \ .
\ee
In the FRW background~(\ref{eq:2.6}), we find
\be
\label{CA3}
F=0\ ,\quad G=24\left(\dot H H^2 + H^4\right)\ .
\ee
For $N$ real scalar, $N_{1/2}$ Dirac spinor, $N_1$ vector fields, $N_2$ ($=0$ or $1$) 
gravitons and $N_{\rm HD}$ higher derivative conformal scalars, 
\bea
\label{CA4}
b &=& \frac{N +6N_{1/2}+12N_1 + 611 N_2 - 8N_{\rm HD}}{120(4\pi)^2}\ ,\nn 
b' &=& - \frac{N+11N_{1/2}+62N_1 + 1411 N_2 -28 N_{\rm HD}}{360(4\pi)^2}\ .
\eea
$b''$ can be arbitrary and we may choose, for example, $b''= - 2b/3$ or $b''=0$. 
If we assume $T_A$ can be given by the effective energy density $\rho_A$ and pressure $p_A$ 
from the conformal anomaly as
\be
\label{CA5}
T_A = - \rho_A + 3 p_A\ ,
\ee
and $\rho_A$ and $p_A$ satisfy the conservation law, 
\be
\label{CA6}
\dot \rho_A + 3H \left( \rho_A + p_A \right) =0\ , 
\ee
we find
\be
\label{CA7}
\rho_A = - \frac{1}{a^4} \int dt a^4 H T_A\ ,\quad 
p_A =  - \frac{1}{3 a^4} \int dt a^4 H T_A + \frac{T_A}{3}\ .
\ee

On the other hand, Eqs.~(\ref{eq:3.10}) and (\ref{eq:3.11}) 
give the effective energy density $\rho_F$ and pressure $p_f$ 
from $f(R)=F(R) - R$ term:
\bea
\label{CA8}
\kappa^2 \rho_F &=&
-\frac{1}{2}\left( F(R)-R \right) +
3\left(H^2  + \dot H\right) \left( F^{\prime}(R)-1 \right)
-18 \left(4H^2 \dot H + H \ddot H\right) F^{\prime\prime}(R)\,, \nn
\kappa^2 p_F &=&
\frac{1}{2}\left( F(R)-R \right)
-\left(3H^2 + \dot H \right) \left( F^{\prime}(R)-1 \right) \nn
&& + 6 \left(8H^2 \dot H + 4{\dot H}^2
+ 6 H \ddot H + \dddot H \right)F^{\prime\prime}(R) +
36\left(4H\dot H + \ddot H\right)^2 F^{\prime\prime\prime}(R)\,.
\eea

We now investigate the magnitude of $\rho_A$, $p_A$, $\rho_F$, and $p_F$ when the phantom 
crossing occurs, when $\dot H=0$. We assume the magnitude of the Hubble rate $H$ could be 
the order of the present Hubble constant $H_0$:
\be
\label{CA9}
H \sim H_0 \sim 10^{-33}\,{\rm eV}\ .
\ee
Expressions on (\ref{CA7}) tells that we may assume $\rho_A \sim p_A \sim T_A$. 
Then we find 
\be
\label{CA10}
\rho_A \sim p_A \sim C H_0^4\ .
\ee
Here $C$ is a dimensionless constant coming from $b$, $b'$, $b''$, and numerical constants and 
therefore $C\sim 10^{2\sim 3}$. 
On the other hand, expressions in (\ref{CA8}) tell that we may assume 
$\rho_F \sim p_F \sim f(R)/\kappa^2$. 
Since $f(R)$ plays the role of the effective cosmological constant, 
we also assume $f(R) \sim H_0^2$ and we find
\be
\label{CA11}
\rho_F \sim p_F \sim \frac{H_0^2}{\kappa^2}\ .
\ee
Since $1/\kappa \sim 10^{28}\,$eV, we find
\be
\label{CA12}
\left| \rho_F \right| \gg \left| \rho_A \right|\ ,\quad 
\left| p_F \right| \gg \left| p_A \right|\ .
\ee
Therefore the quantum correction could be small when the phantom crossing 
occurs and the obtained solutions of the phantom crossing in this paper could 
be stable under the quantum correction. We should note that the quantum 
correction becomes important near the Big Rip singularity, where the curvature 
becomes very large.

\section{Conclusion}

In the present paper, we have considered a crossing of the phantom divide
in modified gravity.
We have reconstructed an explicit model of modified gravity in which
a crossing of the phantom divide can occur
by using the reconstruction method proposed in Ref.~\cite{Nojiri:2006gh}.
As a result, we have shown that the (finite-time) Big Rip singularity appears
in the reconstructed model of modified gravity (i.e., in the Jordan
frame), whereas that in the corresponding scalar field theory obtained through
the conformal transformation (i.e., in the Einstein frame) the singularity 
becomes the infinite-time one.
Furthermore, we have investigated the relations between scalar field theories 
with realizing a crossing of the phantom divide and the corresponding 
modified gravitational ones by using the inverse conformal transformation of 
scalar field theories. 
It has been demonstrated that the scalar field theories describing the
non-phantom phase (phantom one with the Big Rip singularity)
can be represented as the theories of real (complex) $F(R)$ gravity through
the inverse (complex) conformal transformation. 
Moreover, 
taking into account the fact that in the viable
models~\cite{Hu:2007nk,acc,acc1}, which are very realistic modified gravities
that evade solar-system tests, our universe is asymptotically de Sitter space,
we have also proposed a model of modified gravity in which 
the transition from the de Sitter universe to the phantom phase can occur. 
We have found that to construct a viable model generating the transition from
de Sitter space to the phantom phase, additional non-local term or
almost equivalent scalar field is necessary.
It would be interesting to reconsider this problem in the presence of ideal
fluid matter. 
In addition, we have examined the stability for the obtained solutions of the 
crossing of the phantom divide under a quantum correction coming from 
conformal anomaly. 

The study of the future evolution of the universe as discussed in this 
paper may be important to understand whether our universe evolves eternally 
or it will enter into the finite-time singularity. Additionally, this may
shed extra light to some specific properties of different dark energy
models and may help in selecting correct descriptions for the dark side of 
the universe. 

The originality of this work is to reconstruct an explicit model of modified 
gravity in which a crossing of the phantom divide can be realized. 
This is the point beyond the already existing literature. 
We have demonstrated that in principle the crossing of the 
phantom divide can occur at the present time or in the near future in the 
framework of modified gravity without introducing any extra scalar components 
with the wrong kinetic sign such as a phantom. 
This corresponds to the proof that the crossing of the phantom divide 
is possible also for modified gravity theories, which seem to be much less 
pathologic than the usual phantom scalar models of dark energy, 
similar to the scalar field theories in the framework of general relativity. 
The demonstration in this work can be regarded as a significant step to 
construct a more realistic model of modified gravity to 
correctly describe the expansion history of the universe. 

\section*{Acknowledgments}
The work by K.B. and C.Q.G. is supported in part by
the National Science Council of R.O.C. under:
Grant \#s: NSC-95-2112-M-007-059-MY3 and
National Tsing Hua University under Grant \#:
97N2309F1 (NTHU),
that
by S.D.O. was
supported in part by MEC (Spain) projects FIS2006-02842 and
PIE2007-50I023, RFBR grant 06-01-00609 and LRSS project N.2553.2008.2,
and
that by S.N. is supported in part by the Ministry of Education,
Science, Sports and Culture of Japan under grant no.18549001 and Global
COE Program of Nagoya University provided by the Japan Society
for the Promotion of Science (G07).

\appendix
\section{Note on the reconstruction method}

In this appendix, we note the following point on the reconstruction method 
explained in Sec.~II A. 

In the action (\ref{eq:2.2}), if we redefine the auxiliary scalar field $\phi$
by $\phi=\Phi(\varphi)$ by using a proper function $\Phi$, and define
$\tilde P(\varphi)\equiv P(\Phi(\varphi))$ and
$\tilde Q(\varphi)\equiv Q(\Phi(\varphi))$, the action
\begin{eqnarray}
S=\int d^4 x \sqrt{-g}\left\{\frac{1}{2\kappa^2}\left[ \tilde P(\varphi) R
+ \tilde Q(\varphi)\right] + {\cal L}_{\rm matter}\right\}
\label{eq:2A-1}
\end{eqnarray}
is equivalent to the action (\ref{eq:2.2}) because this gives identical $F(R)$
gravity.
This can be explicitly confirmed as follows. First we should note that
Eq.~(\ref{eq:2.3}) is modified as
\begin{eqnarray}
0=\frac{d \tilde P(\varphi)}{d\varphi} R + \frac{d \tilde Q(\varphi)}{d\varphi}
= \left\{ \frac{d P(\Phi(\varphi))}{d\Phi} R + \frac{d Q(\Phi(\varphi))}{d\Phi} \right\}
\frac{d\Phi}{d\varphi}\ .
\label{eq:2A-2}
\end{eqnarray}
Then we can solve $\varphi$ with respect to $R$ by
\begin{eqnarray}
\varphi=\varphi(R) = \Phi^{-1}(\phi(R))\ .
\label{eq:2A-3}
\end{eqnarray}
Here, $\varphi$ is the inverse function of $\Phi$. Hence the obtained
$\tilde F(R)$, corresponding to (\ref{eq:2.4}), is
\begin{eqnarray}
\tilde F(R) \Eqn{\equiv} \tilde P(\varphi(R)) R + \tilde Q(\varphi(R))
= P\left( \Phi\left(\Phi^{-1}(\phi\left(R\right)\right)\right)R
+ Q\left( \Phi\left(\Phi^{-1}(\phi\left(R\right)\right)\right)
\nonumber \\
\Eqn{=} P(\phi(R)) R + Q(\phi(R)) = F(R)\ .
\label{eq:2A-4}
\end{eqnarray}
Thus the obtained $F(R)$ could be identical.
Consequently, there are always ambiguities for the choice in $\phi$ like a
gauge symmetry. In the FRW universe, we now assume that we have solved the
$F(R)$-gravity theory and obtained $R$ as a function of time $t$ as $R=R(t)$.
Then $\phi$ can be expressed as a function of $t$, $\phi=\tilde\phi(t)$, by a
proper function $\tilde\phi$. If we redefine a scalar field by
$\varphi = \tilde\phi^{-1}(\phi)$ by using the inverse function
$\tilde\phi^{-1}$ of $\tilde\phi$, we obtain $\varphi=t$. Hence we can always,
at least locally, identify $\phi$ with time $t$, $\phi=t$, which can be
interpreted as a gauge condition corresponding to the reparameterization of
$\phi=\phi(\varphi)$. 

There could be several cases that we cannot construct $F(R)$.
One possibility is that the differential equation~(\ref{eq:2.11}) has no
consistent solution. Another possibility could be the case that the algebraic
equation (\ref{eq:2.3}) has no solution for obtained $P$ and $Q$
(for example, $P$ and/or $Q$ is a constant).

\section{Reconstruction of an explicit model}

In this appendix, we demonstrate that Eq.~(\ref{PDF2}) can be a solution of 
Eq.~(\ref{eq:2.11}) without matter. 

We start with Eq.~(\ref{eq:2.11}) without matter:
\be
\label{PDF1}
0 = \frac{d^2 P(\phi)}{d\phi^2} - \frac{d \tilde{g}(\phi)}{d\phi}
\frac{d P(\phi)}{d\phi}
+ 2 \frac{d^2 \tilde{g}(\phi)}{d\phi^2} P(\phi)\ .
\ee
By redefining $P(\phi)$ as Eq.~(\ref{PDF2}), 
Eq.~(\ref{PDF1}) is rewritten to
\be
\label{PDF3}
\frac{1}{\tilde{p}(\phi)} \frac{d^2 \tilde{p}(\phi)}{d \phi^2}
= 25 e^{\tilde{g}(\phi)/10} \frac{d^2 \left(e^{-\tilde{g}(\phi)/10}\right)}
{d\phi^2}\ .
\ee
We now consider the model Eq.~(\ref{PDF4}). 
In this case, Eq.~(\ref{PDF3}) is reduced to
\be
\label{PDF5}
\frac{1}{\tilde{p}(\phi)} \frac{d^2 \tilde{p}(\phi)}{d \phi^2}
= \frac{25 \gamma (\gamma + 1)}{\phi^2}\ ,
\ee
which can be solved as Eq.~(\ref{PDF6}). 

We mention the following point about the form of $\tilde{g}(\phi)$
in Eq.~(\ref{PDF4}). From Eq.~(\ref{eq:2.16}), we see that as the universe
evolves, the sign of $\dot{H}$ has to change in time
so that a crossing of the phantom divide can occur.
To realize such a behavior of $H$, there must exist (at least) two terms of
$\phi$ in the brackets $[\,]$ of the logarithmic function on the right-hand
side of Eq.~(\ref{PDF4}). 
(Incidentally, the reason why we select the coefficient `10' on the right-hand 
side of Eq. (2.20) is to obtain the solution analytically.) 
As another form of the Hubble rate realizing a crossing of the phantom divide,
there is the following model~\cite{Nojiri:2005sx}:
\begin{eqnarray}
H = h_0 \left( \frac{1}{t} + \frac{1}{t_s - t} \right)\,,
\label{eq:I9}
\end{eqnarray}
where $h_0$ is a positive constant.
It has been shown that this cosmology can be constructed in terms of
multiple scalar field
theories~\cite{Nojiri:2005sx}.
This form also consists of two terms in $t$.
In this model,
$\dot{H} = h_0 \left(2t-t_s \right)t_s/
\left[t^2 \left(t_s-t \right)^2\right]$.
It follows from Eq.~(\ref{eq:2.16}) that
when $t<t_s/2$, $\dot{H}<0$ and hence
the universe is in non-phantom phase ($w_\mathrm{eff} > -1$),
but that
when $t>t_s/2$, $\dot{H}>0$ and thus
the universe is in phantom phase ($w_\mathrm{eff} < -1$).
It is not hard to formulate the explicit model of modified gravity
with the above type of  the crossing of the phantom divide.

\section{Derivation of the corresponding scalar field theory}

In this appendix, we derive the expression of the action~(\ref{eq:3.8}). 

It follows from the action (\ref{eq:3.1}) that the equation of motion of
the other auxiliary field $\zeta$ is given by
\begin{eqnarray}
\xi = F^{\prime}(\zeta)\,,
\label{eq:3.2}
\end{eqnarray}
where the prime denotes differentiation with respect to $\zeta$.
Substituting Eq.~(\ref{eq:3.2}) into Eq.~(\ref{eq:3.1}) and eliminating
$\xi$ from Eq.~(\ref{eq:3.1}), we find
\begin{eqnarray}
S \Eqn{=}
\int d^{4}x \sqrt{-g}
\left[
\frac{1}{2\kappa^2} \left(
F^{\prime}(\zeta) R + F(\zeta) - F^{\prime}(\zeta) \zeta \right) +
{\mathcal{L}}_{\mathrm{matter}}
\right]\,.
\label{eq:3.3}
\end{eqnarray} 
We make the conformal transformation (\ref{eq:3.4}) with Eq.~(\ref{eq:3.5}) 
of the action~(\ref{eq:3.3}). 
Consequently, the action in the Einstein frame is given by~\cite{F-M, nojiri}
\begin{eqnarray}
S_{\mathrm{E}} =
\int d^{4}x \sqrt{-\hat{g}}
\left[
\frac{1}{2\kappa^2} \left( \hat{R} - \frac{3}{2} \hat{g}^{\mu\nu}
{\partial}_{\mu} \sigma {\partial}_{\nu} \sigma
-V(\sigma) \right)
+ e^{-2\sigma} {\mathcal{L}}_{\mathrm{matter}}
\right]\,,
\label{eq:3.6}
\end{eqnarray}
where
\begin{eqnarray}
V(\sigma) = e^{-\sigma} \zeta(\sigma) - e^{-2\sigma}
F\left( \zeta (\sigma) \right)
= \frac{\zeta}{F^{\prime}(\zeta)} -
\frac{F(\zeta)}{\left( F^{\prime}(\zeta) \right)^2}\,,
\label{eq:3.7}
\end{eqnarray}
and $\hat{g}$ is the determinant of $\hat{g}^{\mu\nu}$.
In deriving Eqs.~(\ref{eq:3.6}) and (\ref{eq:3.7}), we have used
Eq.~(\ref{eq:3.5}).
In addition, $\zeta(\sigma)$ in Eq.~(\ref{eq:3.7}) is obtained by solving
Eq.~(\ref{eq:3.5}) with respect to $\zeta$ as $\zeta=\zeta(\varphi)$. 
By defining $\varphi$ as $\varphi \equiv \sqrt{3/2} \sigma/\kappa$,
the action (\ref{eq:3.6}) is reduced to the form of the canonical
scalar field theory~(\ref{eq:3.8}). 

 From the action (\ref{eq:2.1}), we find that the gravitational field 
equation is given by 
\begin{eqnarray}
F^{\prime}(R) R_{\mu \nu}
- \frac{1}{2}g_{\mu \nu} F(R) + g_{\mu \nu}
\Box F^{\prime}(R) - {\nabla}_{\mu} {\nabla}_{\nu} F^{\prime}(R)
= \kappa^2 T^{(\mathrm{matter})}_{\mu \nu}\,.
\label{eq:3.9}
\end{eqnarray}

When there is no matter,
in the FRW background (\ref{eq:2.6}) the $(\mu,\nu)=(0,0)$ component and
the trace part of the $(\mu,\nu)=(i,j)$ component of Eq.~(\ref{eq:3.9}),
where $i$ and $j$ run from $1$ to $3$, are given by
\begin{eqnarray}
3H^2 =
-\frac{1}{2}\left( F(R)-R \right) +
3\left(H^2  + \dot H\right) \left( F^{\prime}(R)-1 \right)
-18 \left(4H^2 \dot H + H \ddot H\right) F^{\prime\prime}(R)\,,
\label{eq:3.10}
\end{eqnarray}
and
\begin{eqnarray}
&& \hspace{-15mm}
-\left(2\dot H + 3H^2\right) =
\frac{1}{2}\left( F(R)-R \right)
-\left(3H^2 + \dot H \right) \left( F^{\prime}(R)-1 \right) \nonumber \\
&& \hspace{15mm}
{}+ 6 \left(8H^2 \dot H + 4{\dot H}^2
+ 6 H \ddot H + \dddot H \right)F^{\prime\prime}(R) +
36\left(4H\dot H + \ddot H\right)^2 F^{\prime\prime\prime}(R)\,,
\label{eq:3.11}
\end{eqnarray}
respectively.
Using Eqs.~(\ref{eq:3.10}) and (\ref{eq:3.11}), it follows
\begin{eqnarray}
\hspace{-5mm}
2\dot{H}F^{\prime}(R)
+6\left( -4H^2 \dot{H} +4\dot{H}^2 + 3H\ddot{H} + \dddot{H} \right)
F^{\prime\prime}(R)
+36\left(4H\dot H + \ddot H\right)^2 F^{\prime\prime\prime}(R) = 0\,.
\label{eq:3.12}
\end{eqnarray}

\section{Correspondence between the scalar field theories and $F(R)$ gravities 
}

In this appendix, we explore the scalar field theories with realizing a 
crossing of the phantom divide in the Einstein frame 
and consider the behavior of the corresponding theories of $F(R)$ gravity. 

\subsection{Scalar field theories}
The action of scalar field theories in the Einstein frame is given by
\begin{eqnarray}
S_{\Phi} =
\int d^{4}x \sqrt{-\hat{g}}
\left[
\frac{\hat{R}}{2\kappa^2} - \frac{1}{2} \omega \left( \Phi \right)
\hat{g}^{\mu\nu} {\partial}_{\mu} \Phi {\partial}_{\nu} \Phi
-W\left(\Phi\right) \right]\,,
\label{eq:4.1}
\end{eqnarray}
where $\omega \left( \Phi \right)$ is a functions of the scalar field $\Phi$
and $W\left(\Phi\right)$ is the potential of $\Phi$.

In the FRW background (\ref{eq:2.6}), the Einstein equations are given by
\begin{eqnarray}
\frac{3}{\kappa^2} H^2 = \rho_{\Phi}\,,
\quad
-\frac{2}{\kappa^2} \dot{H} = p_{\Phi} + \rho_{\Phi}\,,
\label{eq:4.2}
\end{eqnarray}
where the energy density $\rho_{\Phi}$ of the scalar field $\Phi$ and
the pressure $p_{\Phi}$ of it are given by
\begin{eqnarray}
\rho_{\Phi} = \frac{1}{2} \omega \left( \Phi \right) \dot{\Phi}^2
+W\left(\Phi\right)\,,
\quad
p_{\Phi} = \frac{1}{2} \omega \left( \Phi \right) \dot{\Phi}^2
-W\left(\Phi\right)\,,
\label{eq:4.3}
\end{eqnarray}
respectively. Using equations in (\ref{eq:4.2}) and (\ref{eq:4.3}), we obtain
\begin{eqnarray}
\omega \left( \Phi \right) \dot{\Phi}^2 \Eqn{=} -\frac{2}{\kappa^2}
\dot{H}\,,
\label{eq:4.4} \\
W\left(\Phi\right) \Eqn{=} \frac{1}{\kappa^2}
\left( 3H^2 + \dot{H} \right)\,.
\label{eq:4.5}
\end{eqnarray}

It is the interesting case that
$\omega \left( \Phi \right)$ and $W\left(\Phi\right)$ are defined
in terms of a single function $I\left(\Phi\right)$
as~\cite{Capozziello:2005mj}
\begin{eqnarray}
\omega \left( \Phi \right) \Eqn{=} -\frac{2}{\kappa^2}
\frac{dI\left(\Phi\right)}{d \Phi}\,,
\label{eq:4.6} \\
W\left(\Phi\right) \Eqn{=} \frac{1}{\kappa^2}
\left( 3I^2\left( \Phi \right) + \frac{dI\left(\Phi\right)}{d \Phi}
\right)\,.
\label{eq:4.7}
\end{eqnarray}
Thus we can find the solutions
\begin{eqnarray}
\Phi = t\,,
\quad
H = I(t)\,.
\label{eq:4.8}
\end{eqnarray}
In what follows, we consider the case in which these solutions are satisfied.

If we define a new scalar field $\chi$ as
\begin{eqnarray}
\chi \equiv \int d\Phi \sqrt{|\omega \left( \Phi \right)|}\,,
\label{eq:4.9}
\end{eqnarray}
the action (\ref{eq:4.1}) can be rewritten to the form in Eq.~(\ref{eq:4.10}), 
where the sign in front of the kinetic term depends on that of
$\omega \left( \Phi \right)$. If the sign of $\omega \left( \Phi \right)$
is positive (negative), that of the kinetic term is $-$ ($+$).
In the non-phantom phase, the sign of the kinetic term is always $-$, and
in the phantom one it is always $+$.
In principle, it follows from Eq.~(\ref{eq:4.9}) that $\Phi$ can be solved
with respect to $\chi$ as $\Phi = \Phi (\chi)$. Hence, the potential
$\tilde{W}(\chi)$ is given by
$\tilde{W}(\chi) = W \left( \Phi (\chi) \right)$.

In the case of the model explained in Sec.~II B, it follows from
Eq.~(\ref{PDF9}) that $I\left( \Phi \right)$ is given by
\begin{eqnarray}
I\left( \Phi \right)
= \left( \frac{10}{\Phi} \right)
\left[
\frac{\gamma + \left( \gamma + 1 \right)
\left( \frac{\Phi}{t_s} \right)^{2\gamma+1}}
{1- \left( \frac{\Phi}{t_s} \right)^{2\gamma+1}} \right]\,.
\label{eq:4.11}
\end{eqnarray}
  From the solutions in (\ref{eq:4.8}), we find that Eq.~(\ref{eq:4.11}) gives
\begin{eqnarray}
H =
\left( \frac{10}{t} \right)
\left[
\frac{\gamma + \left( \gamma + 1 \right)
\left( \frac{t}{t_s} \right)^{2\gamma+1}}
{1- \left( \frac{t}{t_s} \right)^{2\gamma+1}} \right]\,.
\label{eq:4.12}
\end{eqnarray}
In deriving the expressions in (\ref{eq:4.11}) and (\ref{eq:4.12}),
we have used Eq.~(\ref{PDF8}).
In this case, from Eqs.~(\ref{eq:4.6}), (\ref{eq:4.7}) and (\ref{eq:4.9})
we find
\begin{eqnarray}
\hspace{-10mm}
\omega \left( \Phi \right) \Eqn{=}
\frac{
20 \left[
\gamma - 4\gamma \left( \gamma+1 \right)
\left( \frac{\Phi}{t_s} \right)^{2\gamma+1} -
\left( \gamma+1 \right) \left( \frac{\Phi}{t_s}
\right)^{2\left( 2\gamma+1 \right)} \right]}
{\kappa^2 \Phi^2
\left[ 1-\left( \frac{\Phi}{t_s} \right)^{2\gamma+1} \right]^2}\,,
\label{eq:4.13} \\
\hspace{-10mm}
\chi \Eqn{=} \frac{\sqrt{20}}{\kappa} \int d\Phi
\frac{\sqrt{\left|
\gamma - 4\gamma \left( \gamma+1 \right)
\left( \frac{\Phi}{t_s} \right)^{2\gamma+1} -
\left( \gamma+1 \right) \left( \frac{\Phi}{t_s}
\right)^{2\left( 2\gamma+1 \right)}
\right|}}
{\Phi \left[ 1-\left( \frac{\Phi}{t_s} \right)^{2\gamma+1} \right]}\,,
\label{eq:4.14} \\
\hspace{-10mm}
\tilde{W}(\chi) \Eqn{=}
\frac{10}
{\kappa^2 \Phi^2 (\chi)
\left[ 1-\left( \frac{\Phi(\chi)}{t_s} \right)^{2\gamma+1} \right]^2}
\nonumber \\
&&
\hspace{-12mm}
{}\times
\left[
\gamma \left( 30\gamma -1 \right) +64\gamma \left( \gamma+1 \right)
\left( \frac{\Phi(\chi)}{t_s} \right)^{2\gamma+1}
+ \left( \gamma+1 \right) \left( 30\gamma+31 \right)
\left( \frac{\Phi(\chi)}{t_s} \right)^{2\left( 2\gamma+1 \right)} \right]
\,.
\label{eq:4.15}
\end{eqnarray}

As shown in Sec.~II B, when $t<t_\mathrm{c}$, $\dot{H}<0$ (non-phantom phase)
and it follows from Eq.~(\ref{eq:4.4}) that $\omega >0$.
In the non-phantom phase,
the sign of the kinetic term in the action (\ref{eq:4.10}) is $-$.
On the other hand, when $t>t_\mathrm{c}$,
$\dot{H}>0$ (phantom phase) and from Eq.~(\ref{eq:4.4}) we see that
$\omega <0$. In the phantom phase,
the sign of the kinetic term in the action (\ref{eq:4.10}) is $+$.
When $t=t_\mathrm{c}$, $\omega =0$ and the transition from non-phantom phase
to phantom one occurs~\cite{Nojiri:2005pu}.

\subsection{Corresponding theories of $F(R)$ gravity}
We investigate the behavior of the modified gravity (\ref{eq:4.25}) 
in the limit of $t \to t_s$ 
for the case in which the potential $\tilde{W}\left(\chi(R)\right)$ is given
by Eq.~(\ref{eq:4.15}). 
If the scalar field $\chi$ is real,
it follows from Eq.~(\ref{eq:4.23}) that the scalar curvature $R$ is
not always real.
In order for $R$ to be real, the following condition
should be satisfied~\cite{Briscese:2006xu}:
\begin{eqnarray}
e^{i \sqrt{2/3} \kappa \chi}
\left( 4\kappa^2 \tilde{W}(\chi) - i \sqrt{6} \kappa
\frac{d \tilde{W}(\chi)}{d \chi} \right)
= e^{-i \sqrt{2/3} \kappa \chi}
\left( 4\kappa^2 \tilde{W}(\chi) + i \sqrt{6} \kappa
\frac{d \tilde{W}(\chi)}{d \chi} \right)\,.
\label{eq:4.26}
\end{eqnarray}
This condition is reduced to
\begin{eqnarray}
\frac{1}{\tilde{W}(\chi)} \frac{d \tilde{W}(\chi)}{d \chi} =
2\sqrt{\frac{2}{3}} \kappa
\tan \sqrt{\frac{2}{3}} \kappa \chi \,.
\label{eq:4.27}
\end{eqnarray}
Except for the form of $\tilde{W}(\chi)$ satisfying Eq.~(\ref{eq:4.27}),
$R$ is complex if $\chi$ is real.
The form of $\tilde{W}(\chi)$ in Eq.~(\ref{eq:4.15}) cannot
satisfy Eq.~(\ref{eq:4.27}).
This applies to the case in which the Hubble rate is given by
Eq.~(\ref{eq:I9}).
If the scalar field $\chi$ is pure imaginary and expressed as $\chi = i \eta$,
where $\eta$ is a real scalar field, and
the potential $\tilde{W}(\chi)$ contains only even power of $\chi$,
Eq.~(\ref{eq:4.23}) is rewritten to
\begin{eqnarray}
R= e^{\mp \sqrt{2/3} \kappa \eta}
\left[ 4\kappa^2 \tilde{W}(-\eta^2) \pm 2\sqrt{6} \kappa
\frac{d \tilde{W}(-\eta^2)}{d \left( -\eta^2 \right)} \right]\,,
\label{eq:4.28}
\end{eqnarray}
which tells that $R$ is real.
In fact, however, that the action (\ref{eq:4.10})
with the $+$ sign of the kinetic term is reduced to~\cite{Briscese:2006xu}
\begin{eqnarray}
S_{\eta} =
\int d^{4}x \sqrt{-\hat{g}}
\left[
\frac{\hat{R}}{2\kappa^2} - \frac{1}{2}
\hat{g}^{\mu\nu} {\partial}_{\mu} \eta {\partial}_{\nu} \eta
-\tilde{W}(-\eta^2) \right]\,.
\label{eq:4.29}
\end{eqnarray}
This action corresponds to the non-phantom (canonical) theory.
As a consequence, if the scalar field theory describes the phantom phase with
the Big Rip singularity, the corresponding theory of $F(R)$ gravity
is usually complex. (Note that counterexample is
known~\cite{Briscese:2006xu}, but
even in this case, when $t \to t_s$, the modified gravity becomes complex.)
When $t \to t_s$, from Eq.~(\ref{eq:4.15})
$\tilde{W}(\chi)$ diverges and hence $F_\mathrm{P}(R)$ also diverges. 

In the Einstein frame, the scalar field couples with matter 
and therefore, the frame could be unphysical. The coupling changes the scale of time interval. We usually measure the time by using electromagnetism. In the 
limit of neglecting the local gravity, the measured time corresponds to the 
cosmological time in the Jordan frame but not in the Einstein frame.

\section{Examples of viable $F(R)$ gravity models}

In this appendix, we show examples of viable $F(R)$ gravity models. 

The modified part in $F(R)$ in the action (\ref{eq:2.1}) can be separated as
\be
\label{fr2}
F(R) = R + f(R)\ ,
\ee
where $f(R)$ is an arbitrary function of the scalar curvature $R$.
The interesting viable model proposed in Ref.~\cite{Hu:2007nk} is given by 
\be
\label{HS1}
f(R) = f_\mathrm{HS}(R) \equiv
-\frac{\bar{M}^2 c_2 \left(R/\bar{M}^2\right)^p}{c_3
\left(R/\bar{M}^2\right)^p + 1}
= - \frac{\bar{M}^2 c_2}{c_3}
+ \frac{\bar{M}^2 c_2/c_3}{c_3 \left(R/\bar{M}^2\right)^p + 1}\ ,
\ee
where $c_2$ and $c_3$ are dimensionless constants, $p$ is a positive
constant, and $\bar{M}$ denotes a mass scale. 
In this model, when $R/\bar{M}^2 \to \infty$,
$f_\mathrm{HS}(R) \sim - \bar{M}^2 c_2/c_3 = \mathrm{const}$. 

The generalizations of the above model which admits the unification 
of early-time inflation with late-time acceleration (with intermediate 
radiation/matter dominance) have been proposed in Ref.~\cite{acc}. 
The example of such a theory is given by 
\begin{eqnarray}
f(R) \Eqn{=} -\alpha_0 \left( \tanh \left( \frac{b_0 \left(R-R_0\right)}{2}
\right) + \tanh \left( \frac{b_0 R_0}{2} \right) \right) \nonumber \\
&&
{}-\alpha_I \left( \tanh \left( \frac{b_I \left(R-R_I\right)}{2} \right)
+ \tanh \left( \frac{b_I R_I}{2} \right) \right)\,,
\label{eq:5J-0-0}
\end{eqnarray}
where $\alpha_0$, $\alpha_I$, $b_0$ and $b_I$ are constant, and $R_0$ and 
$R_I$ are constant scalar curvatures. 
We note that such a theory may correctly describe the whole expansion 
history of the universe qualitatively: inflation, radiation/matter dominance 
and dark energy, as it has been explained in Ref.~\cite{nojiri}.

\section{Numerical calculations with the matter}

In this appendix, for the case Eq.~(\ref{dp1}) with $t_s g_0 = g_1$, 
we explain the numerical calculations of Eq.~(\ref{eq:2.11}) with the 
matter in detail.

We define $\tilde{g}_0 \equiv t_s g_0$ and $Y \equiv 1- X$ with 
$X \equiv t/t_s$. 
From Eqs.~(\ref{eq:2.10}) and (\ref{dp1}) with $t_s g_0 = g_1$, 
$H= d \tilde{g}(t)/\left(d t \right)$ and $R=6\left( \dot{H} + 2H^2 \right)$, 
we obtain 
\begin{eqnarray}
\tilde{g}(t) \Eqn{=} -\tilde{g}_0 \left( Y+\log Y \right)\,, 
\label{eq:F.1} \\
a(t) \Eqn{=} \left[ \left( 1-\frac{1}{\alpha} \right) \frac{1}{Y} 
\right]^{\tilde{g}_0} 
\exp \left[ \tilde{g}_0 \left( 1-\frac{1}{\alpha}-Y \right) \right]\,,
\label{eq:F.2} \\
\tilde{R} \Eqn{=} t_s^2 R = 6 \tilde{g}_0 \left( 2\tilde{g}_0 + 
\frac{4\tilde{g}_0}{Y} + \frac{2\tilde{g}_0 +1}{Y^2} \right)\,,
\label{eq:F.3}
\end{eqnarray}
where we have taken $\bar{a}= \left(1-1/\alpha \right)^{\tilde{g}_0} 
\exp \left[ \tilde{g}_0 \left(1-1/\alpha \right) \right]$ 
so that the present value of the scale factor should be unity.

The solution of Eq.~(\ref{eq:F.3}) with respect to $Y$ is given by 
\begin{eqnarray}
Y (\tilde{R}) = \frac{1}{2\tilde{g}_0^2 -\tilde{R}/6} 
\left[ -2\tilde{g}_0^2 \pm \sqrt{-2\tilde{g}_0^3 + 
\frac{\tilde{g}_0 \left(2\tilde{g}_0 +1\right)\tilde{R}}{6}}
\right]\,. 
\label{eq:F.4}
\end{eqnarray}
In what follows, we use the lower sign in Eq.~(\ref{eq:F.4}). 

For simplicity, we consider the case in which there exists a matter with a 
constant EoS parameter $w=p/\rho$. In this case, 
by using Eq.~(\ref{eq:F.4}), Eqs.~(\ref{eq:2.11}) and (\ref{eq:2.12}) 
are rewritten to 
\begin{eqnarray}
&&
\left[ 12\tilde{g}_0 \left( 2\tilde{g}_0 Y + 2\tilde{g}_0 +1 \right) 
\right]^2 \frac{d^2 P(\tilde{R})}{d^2 \tilde{R}} 
\nonumber \\
&&
{}
+\tilde{g}_0 Y^2 \left\{ 12 \left[ 4\tilde{g}_0 Y +3\left( 2\tilde{g}_0 +1
\right) \right] +Y^3\left(1+Y\right) \right\}
\frac{d P(\tilde{R})}{d \tilde{R}} 
{}+2\tilde{g}_0 Y^4 P(\tilde{R}) 
\nonumber \\
&& 
{}+ Y^6 t_s^2 \kappa^2 \left(1+w\right) \bar{\rho} 
\left\{
\left[ \left( 1-\frac{1}{\alpha} \right) \frac{1}{Y} 
\right]^{\tilde{g}_0} 
\exp \left[ \tilde{g}_0 \left( 1-\frac{1}{\alpha}-Y \right) \right]
\right\}^{-3\left(1+w\right)} =0
\label{eq:F.5}
\end{eqnarray}
and
\begin{eqnarray}
t_s^2 Q(\tilde{R}) \Eqn{=} 
-6\tilde{g}_0^2 \left( \frac{1+Y}{Y}\right)^2 P(\tilde{R}) 
-72\tilde{g}_0^2 \left( \frac{1+Y}{Y^3}\right) 
\left( 2\tilde{g}_0 + \frac{2\tilde{g}_0 +1}{Y}\right) 
\frac{d P(\tilde{R})}{d \tilde{R}} 
\nonumber \\
&&
{}+2 t_s^2 \kappa^2 \bar{\rho} 
\left\{
\left[ \left( 1-\frac{1}{\alpha} \right) \frac{1}{Y} 
\right]^{\tilde{g}_0} 
\exp \left[ \tilde{g}_0 \left( 1-\frac{1}{\alpha}-Y \right) \right]
\right\}^{-3\left(1+w\right)}\,,
\label{eq:F.6}
\end{eqnarray}
respectively. 
Here, $\bar{\rho}$ corresponds to the 
present energy density of the matter. In particular, 
we use the present value of the 
cold dark matter with $w=0$ for $\bar{\rho}$, i.e., 
$\bar{\rho} = 0.233 \rho_\mathrm{c}$~\cite{Komatsu:2008hk}, 
where $\rho_\mathrm{c} =3H_0^2/\left(8 \pi G \right) =
3.97 \times 10^{-47} \mathrm{GeV}^4$ is the critical energy density. From 
Eq.~(\ref{eq:2.4}) and $\tilde{R} = t_s^2 R$, we have 
\begin{eqnarray}
\frac{F(\tilde{R})}{2\kappa^2} = \frac{1}{2\kappa^2 t_s^2}
\left( P(\tilde{R}) \tilde{R} + t_s^2 Q(\tilde{R}) \right)\,.
\label{eq:F.7}
\end{eqnarray}
To examine $F(\tilde{R})$, we numerically solve 
Eqs.~(\ref{eq:F.5})--(\ref{eq:F.7}).

\begin{figure}[tbp]
\begin{center}
   \includegraphics{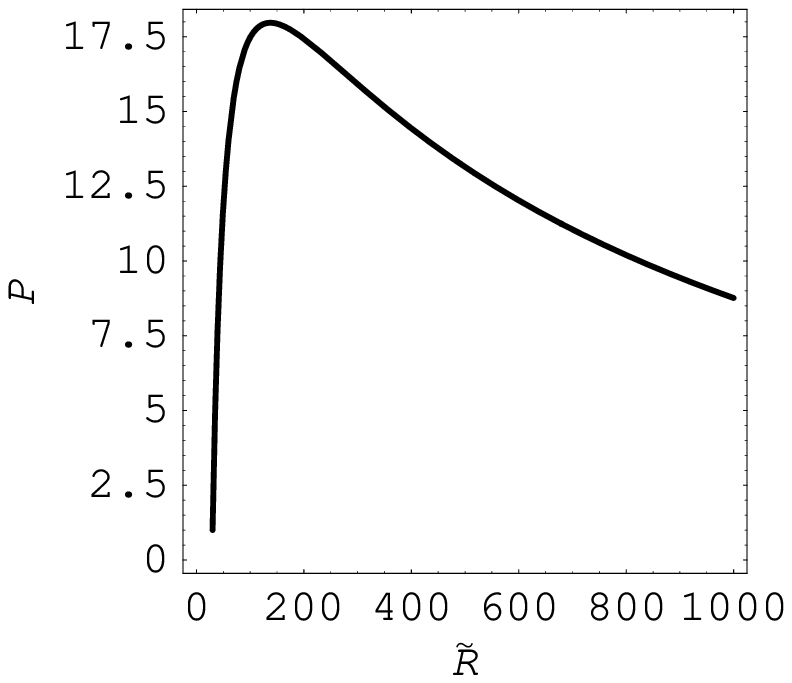}
   \includegraphics{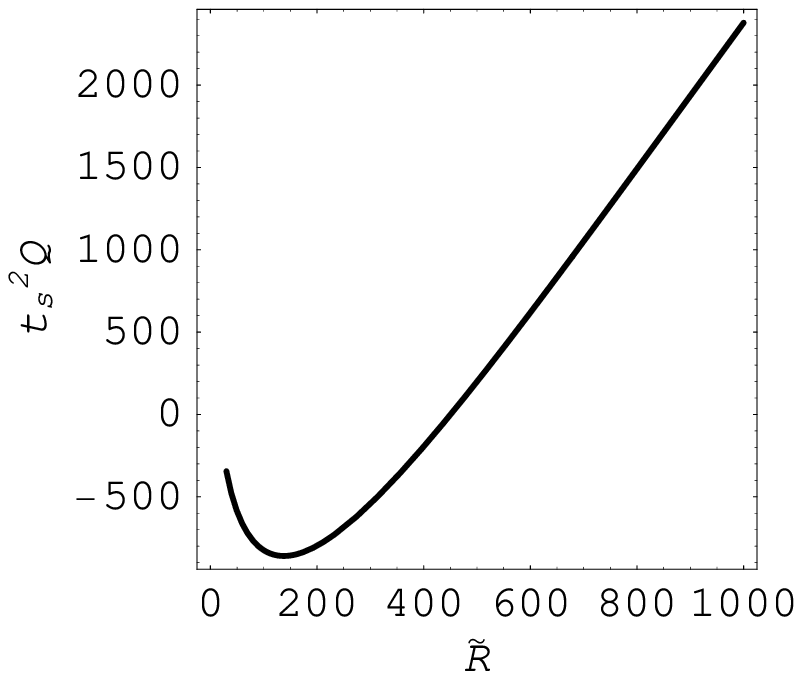}
\caption{
$P(\tilde{R})$ and $Q(\tilde{R})/\nu^2$ as functions of $\tilde{R}$ for 
$t_s = 2t_0$ and $\bar{\rho} = 0.233 \rho_\mathrm{c}$. 
}
\end{center}
\label{fg:3}
\end{figure}

In Fig.~3, we depict $P(\tilde{R})$ and $Q(\tilde{R})/\nu^2$ as 
functions of $\tilde{R}$. 
The range of $\tilde{R}$ is given by $30 \leq \tilde{R} \leq 1000$, 
corresponding to $0 < X =t/t_s < 1$. 
Here, we have taken the initial conditions as $P(\tilde{R} = 30) =1.0$ and 
$d P(\tilde{R} = 30)/ ( d \tilde{R} ) =1.0$. 
By using Eq.~(\ref{eq:F.7}), 
we show the behavior of $F(\tilde{R})/\left(2 \kappa^2 \right)$ in Fig.~2.


\end{document}